\documentclass[twocolumn]{aastex7}

\hypersetup{linkcolor=blue,citecolor=blue,filecolor=blue,urlcolor=blue}

\usepackage{epsfig}
\usepackage{epstopdf}  
\usepackage{placeins}
\usepackage{graphicx,color}     
\usepackage{amssymb} 
\usepackage{float}
\usepackage{color}     
\usepackage{url}       
\usepackage{amsmath}
\usepackage{rotating}           
\usepackage{float}            
\usepackage{textcomp}           
\usepackage{dcolumn}
\usepackage{times}
\usepackage{placeins}  % Add this in your preamble if not already included
\usepackage{xcolor}
\usepackage{hyperref}
\usepackage{tabularx}
\usepackage[english]{babel}
\usepackage{amssymb}
\usepackage{pifont}
\usepackage{hyperref}
\usepackage[utf8]{inputenc}
\begin{document}
\shorttitle{Chromospheric Fan-shaped Jets}
\shortauthors{Bura et al.}
\title{Formation of chromospheric fan-shaped jets through magnetic reconnection}

\author[0009-0000-5018-9735]{Annu Bura}
\affiliation{Indian Institute of Astrophysics, Koramangala, Bangalore 560034, India}
\affiliation{Pondicherry University, R.V. Nagar, Kalapet 605014, Puducherry, India}
\email{annu.bura@iiap.res.in}  

\author[orcid=0000-0002-9667-6392]{Tanmoy Samanta} 
\affiliation{Indian Institute of Astrophysics, Koramangala, Bangalore 560034, India}
\affiliation{Pondicherry University, R.V. Nagar, Kalapet 605014, Puducherry, India}
\email[show]{tanmoy.samanta@iiap.res.in}

\author[orcid=0000-0003-0819-464X]{Avijeet Prasad}
\affiliation{Institute of Theoretical Astrophysics, University of Oslo, PO Box 1029 Blindern, 0315, Oslo, Norway}
\affiliation{Rosseland Centre for Solar Physics, University of Oslo, PO Box 1029 Blindern, 0315, Oslo, Norway}
\email{avijeet.ka.id@gmail.com}

\author[orcid=0000-0002-5691-6152]{Ronald L. Moore}
\affiliation{NASA Marshall Space Flight Center, Huntsville, AL 35812, USA}
\affiliation{Center for Space Plasma and Aeronomic Research, University of Alabama in Huntsville, Huntsville, AL 35805, US}
\email{ronald.l.moore@nasa.gov}

\author[orcid=0000-0003-1281-897X]{Alphonse C. Sterling}
\affiliation{NASA Marshall Space Flight Center, Huntsville, AL 35812, USA}
\email{alphonse.sterling@nasa.gov}

\author[orcid=0000-0001-9982-2175]{Vasyl Yurchyshyn}
\affiliation{Big Bear Solar Observatory, New Jersey Institute of Technology, 40386 North Shore Lane, Big Bear, CA 92314, USA}
\email{vasyl.yurchyshyn@njit.edu}

\author[orcid=0000-0002-9967-0391]{Arun Surya}
\affiliation{Indian Institute of Astrophysics, Koramangala, Bangalore 560034, India}
\email{arun.surya@iiap.res.in}

\begin{abstract}
Recurrent chromospheric fan-shaped jets highlight the highly dynamic nature of the solar atmosphere. They have been named as ``light walls" or ``peacock jets" in high-resolution observations. In this study, we examined the underlying mechanisms responsible for the generation of recurrent chromospheric fan-shaped jets utilizing data from the Goode Solar Telescope (GST) at Big Bear Solar Observatory, along with data from the Atmospheric Imaging Assembly (AIA) and the Helioseismic and Magnetic Imager (HMI) onboard the Solar Dynamic Observatory (SDO). These jets appear as dark elongated structures in H$\alpha$ wing images, persist for over an hour, and are located in the intergranular lanes between a pair of same-polarity sunspots. Our analysis reveals that magnetic flux cancellation at the jet base plays a crucial role in their formation. HMI line-of-sight magnetograms show a gradual decrease in opposite-polarity fluxes spanning the sequence of jets in H$\alpha$ - 0.8 \AA\ images, suggesting that recurrent magnetic reconnection, likely driven by recurrent miniature flux-rope eruptions that are built up and triggered by flux cancellation, powers these jets. Additionally, magnetic field extrapolations reveal a 3D magnetic null-point topology at the jet formation site $\sim$1.25 Mm height. Furthermore, we observed strong brightening in AIA 304 \AA\ channel above the neutral line. Based on our observations and extrapolation results, we propose that these recurrent chromospheric fan-shaped jets align with the minifilament eruption model previously proposed for coronal jets. Though our study focuses on fan-shaped jets in between same-polarity sunspots, similar mechanism might be responsible for light bridge-associated fan-shaped jets.
\end{abstract}
\section{Introduction} \label{sec:Intro} 
The solar atmosphere exhibits diverse jetting phenomena, appearing as collimated plasma beams that span a broad range of spatial and temperature scales, from large-scale X-ray jets (\citealt{1992PASJ...44L.173S}; \citealt{1996ApJ...464.1016C}; \citealt{1996PASJ...48..123S}; \citealt{2015Natur.523..437S}) and EUV jets (\citealt{2009SoPh..259...87N};  \citealt{2016ApJ...828L...9S}) in the corona to small-scale jets (\citealt{2007Sci...318.1591S}; \citealt{2011A&A...533A..76K}; \citealt{2014Sci...346A.315T};  \citealt{2022ApJ...938..122P}) in the lower solar atmosphere. These solar jets have been proposed to be often triggered by magnetic flux emergence and/or cancellation processes (\citealt{2007A&A...469..331J}; 
\citealt{2014Sci...346C.315P}; 
\citealt{2015ApJ...815...71C}; \citealt{2017ApJ...844..131P}; \citealt{2018ApJ...869...78C}; \citealt{2019ApJ...887..220Y}; \citealt{2019ApJ...887..239Y}; \citealt{2020ApJ...904...15Y}; \citealt{2019Sci...366..890S}; \citealt{2021ApJ...918L..20H};
\citealt{2022NatCo..13..640Y};
\citealt{2023ApJ...942...86Y}) and can occur in all types of solar regions, including active regions (ARs), coronal holes, and quiet-Sun regions.\\ 
\begin{figure*}[!htbp]
\renewcommand{\thefigure}{1}
\centering
\includegraphics[width=\textwidth]{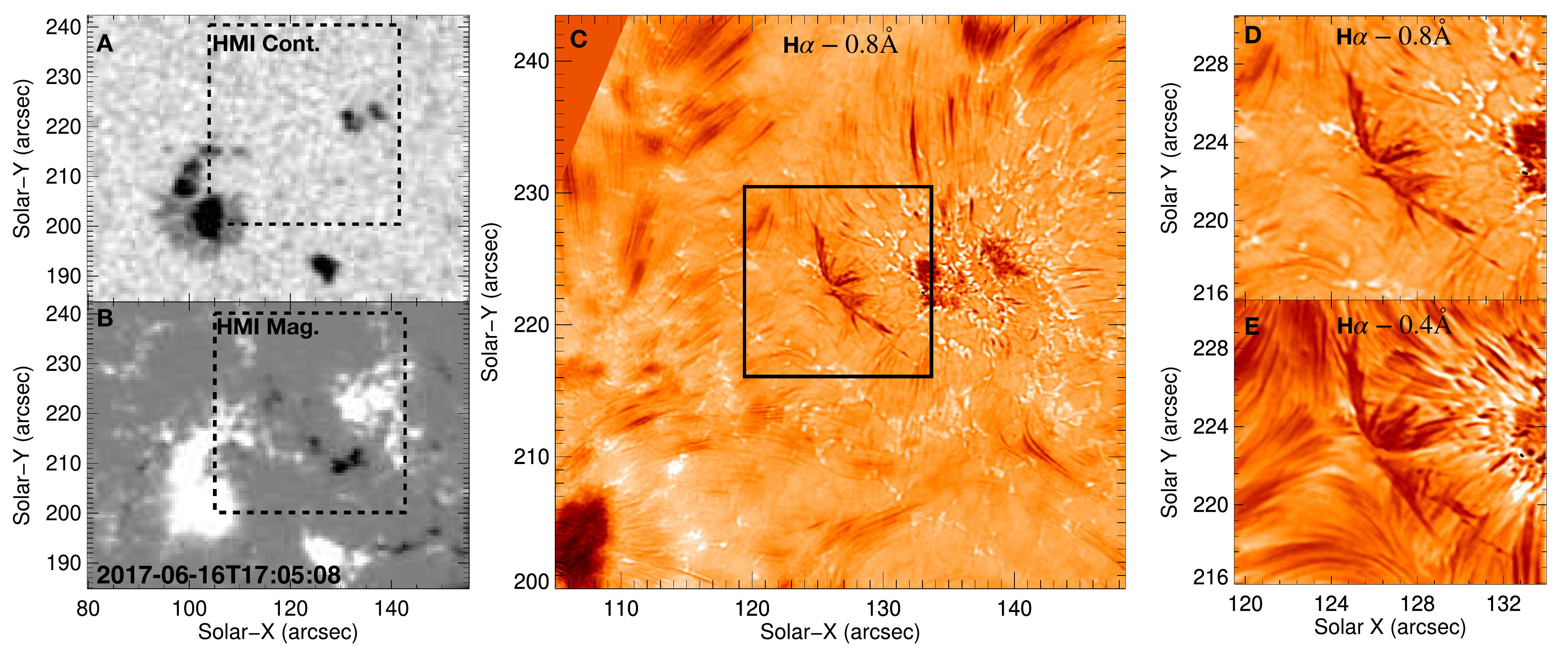}
\caption{Panels A and B show images of HMI continuum and HMI line of sight magnetic field, respectively, taken on 2017 June 16. Panel C shows a blue wing image (at H$\alpha- 0.8$ \AA) observed by GST at the location of the black dotted box of panels  (A, B). Panels (D-E) show a zoomed-in view of the black box region in panel C. The H$\alpha$ - 0.8~\AA\ and H$\alpha$ - 0.4~\AA\ images clearly show fan-shaped jets.  An animation depicting the evolution of panel C from 16:46:10 UT to 20:28:22 UT is available online.}
\label{fig: fig1}
\end{figure*}
\indent Recurrent fan-like jets, which appear as a series of successive eruptions in high-resolution chromospheric observations, highlight the highly dynamic nature of the solar atmosphere. Earlier studies, constrained by spatial resolution, referred to them as H$\alpha$ surges, plasma ejections, or chromospheric jets (\citealt{1973SoPh...32..139R}; \citealt{2001ApJ...555L..65A}; \citealt{2014A&A...567A..96L}). Recent high-resolution observations have named them as light walls (\citealt{2015ApJ...804L..27Y}) or peacock jets (\citealt{2016A&A...590A..57R}). However, their study remains limited due to the rarity of such observations. \\
\indent \citet{1973SoPh...32..139R} first reported these jets as H$\alpha$ surges ejecting from a light bridge (LB) of a sunspot umbra. They reported rapid initial acceleration with peak velocities of 150 km~s$^{-1}$ and extending up to 50 Mm. After an ascending phase, a descending phase was observed with an acceleration smaller than gravity, suggesting the presence of a breaking force. The understanding of these jets remained limited until the advent of high-resolution ground-based observations. Combining ground-based H$\alpha$ and space-based 171 \AA\ observations, \citet{2001ApJ...555L..65A} reported recurrent plasma ejections from a LB in a mature sunspot. The H$\alpha$ surges ($\sim$50 km~s$^{-1}$) appeared as dark recurrent jets, while the 171 \AA\ ejections had a loop-like structure. The absence of ejections in images from the soft X-ray telescope (SXT) aboard Yohkoh (\citealt{1991SoPh..136...37T}) suggests the ejected plasma had temperatures below a few MK. \citet{2001ApJ...555L..65A} found maximum velocities of around 40 km~s$^{-1}$, lengths of 15–20 Mm, and mean lifetimes of 10 minutes. \\
\indent These chromospheric fan-shaped jets are mostly observed above sunspot LBs (\citealt{1973SoPh...32..139R}; \citealt{2001ApJ...555L..65A}; \citealt{2007MNRAS.376.1291B}; \citealt{2015ApJ...804L..27Y};  \citealt{2016A&A...590A..57R}; \citealt{2018ApJ...854...92T}; \citealt{2019ApJ...870...90B}), where the photospheric magnetic field transitions from vertical to horizontal (\citealt{2009ApJ...696L..66S}). It is suggested that the shear between the vertical umbral magnetic field and the horizontal magnetic field of LBs can generate a sharp current layer, creating favorable conditions for magnetic reconnection, which in turn drives the formation of fan-shaped jets (\citealt{2009ApJ...696L..66S}; \citealt{2015ApJ...811..137T}; \citealt{2018A&A...609A..14R}). Recently, fan-shaped jets have been associated with flares, which can influence their dynamics and structure. \citet{2016ApJ...829L..29H} reported that flares along the same field lines as these jets cause the jets to incline and shorten. \citet{2022A&A...659A..58P} found that an expanding flare ribbon suppressed the upward motion of fan-shaped jets by covering their base during an X9.3 flare.\\
\indent Using 3D magnetohydrodynamics simulations, \citet{2011RAA....11..701J} reproduced the structure of fan-shaped jets and showed that they formed through shearing reconnection, with initial acceleration driven by the Lorentz force, followed by gas pressure gradients. \citet{2016ApJ...826..217L} used high-resolution observations with the Goode Solar Telescope (GST; \citealt{2010AN....331..636C}; \citealt{2012SPIE.8444E..03G}) at Big Bear Solar Observatory (BBSO) to show that the gradient in gas pressure is created via the heating at the base of jets.\\
\indent These recurrent jets often appear as dark features in H$\alpha$ observations, with a bright front in the transition region and coronal lines. Their footpoint exhibits brightening in \ion{Ca}{2} H (\citealt{2009ApJ...696L..66S}), H$\alpha$ (\citealt{2016A&A...590A..57R}; \citealt{2020MNRAS.492.2510L}) and \ion{Ca}{2} 8542 \AA (\citealt{2018A&A...609A..14R}) lines, while the front appears bright in images from the Atmospheric Imaging Assembly (AIA; \citealt{2012SoPh..275...17L}) onboard Solar Dynamics Observatory (SDO) - 304 \AA\ and 171 \AA\ channels (\citealt{2016A&A...590A..57R}; \citealt{2018ApJ...855L..19R}), and in images from the Interface Region Imaging Spectrograph (IRIS; \citealt{2014SoPh..289.2733D}) 1330  \AA\ channel (\citealt{2015MNRAS.452L..16B}; \citealt{2015ApJ...804L..27Y}; \citealt{2017ApJ...838....2Z}; \citealt{2018ApJ...854...92T}; \citealt{2019ApJ...870...90B}). Such bright fronts ahead of H$\alpha$ surges have been suggested to result from heating to transition region temperatures due to shock fronts or compression (\citealt{2015MNRAS.452L..16B}), or as oscillations driven by leaked p-mode waves from below the photosphere (\citealt{2015ApJ...804L..27Y}; \citealt{2017ApJ...838....2Z}; \citealt{2018ApJ...855L..19R}).\\
\begin{figure*}[t]
\renewcommand{\thefigure}{2}
\centering
\includegraphics[width=\textwidth]{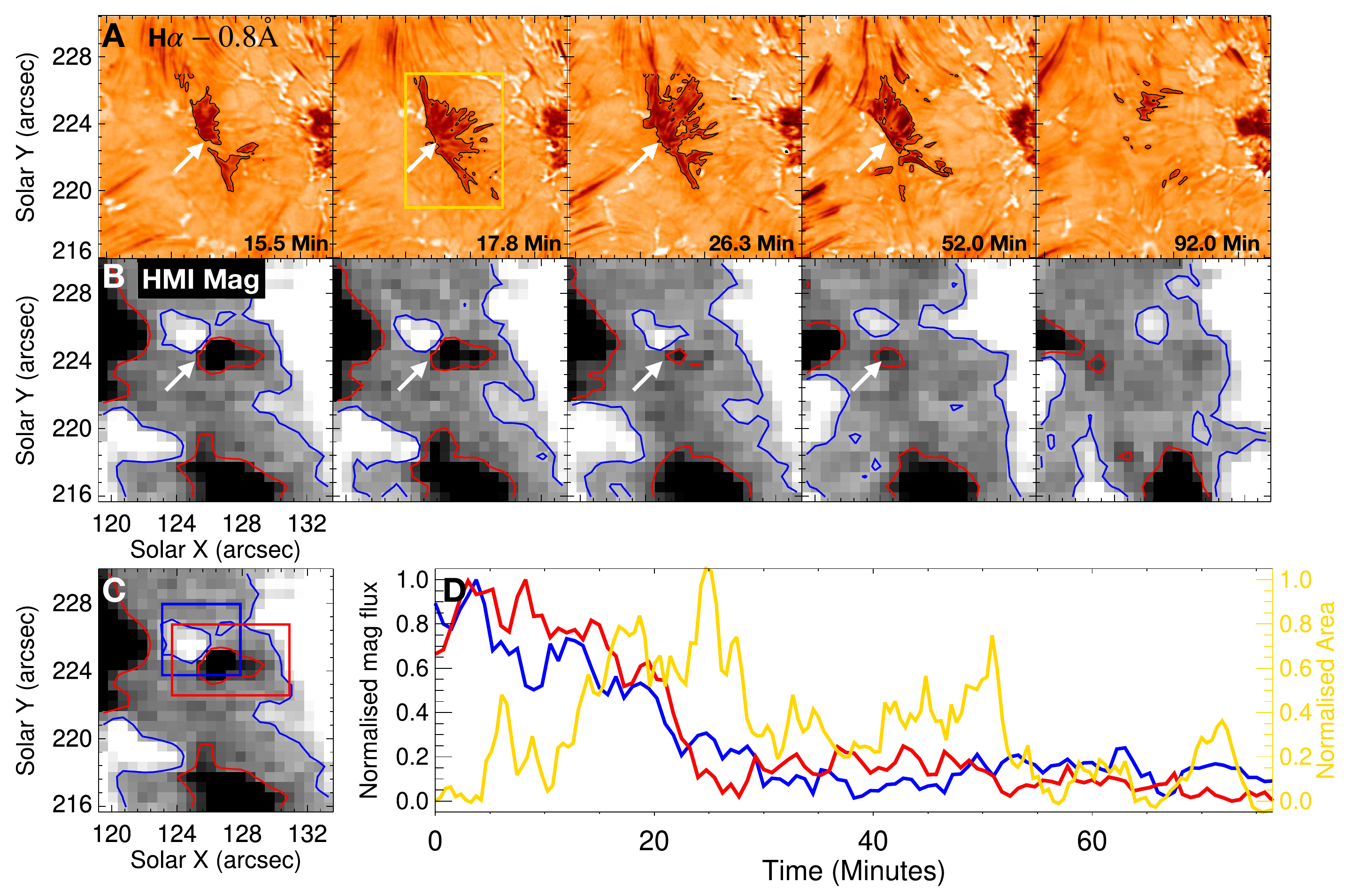}
\caption{Panels of A show the temporal evolution around the fan-shaped jets. The black contours outline the jet area, plotted in panel D. Panels of B show the corresponding magnetic field evolution. The blue and red contours represent the flux above $+20$ and $-20 G$, respectively. The white arrows highlight the base of dark recurrent fan-shaped jets. Panel C is the same as the left-most subpanel of B. The blue and red square boxes that cover the positive and negative polarity of the dipole are used to compute the magnetic flux of the two opposite polarities. D: Variation of measured positive (blue line) and negative flux (red line) over time. The golden curve shows the variation in the area of the fan-shaped jet over time within the rectangular box marked in panel A. The timestamps in panel A show the time in minutes elapsed from 16:46:10 UT.}
\label{fig: fig2}
\end{figure*}
\indent Recurrent fan-shaped jets are observed not only above LBs, but also between distinct sunspots (\citealt{2016A&A...589L...7H}; \citealt{2022ApJ...932...95Z}; \citealt{2024A&A...691A.198J}) and within penumbra (\citealt{1973SoPh...32..139R}; \citealt{2018ApJ...855L..19R}; \citealt{2016A&A...589L...7H}).  Recently, \citet{2020ApJ...904...84L} detected the emergence and cancellation of opposite magnetic polarity fluxes at the base of LB-associated fan-shaped jets. \citet{2022ApJ...932...95Z} observed a decrease in the magnetic flux of minority-polarity fields near the footpoints of fan-shaped jets located in the intergranular lanes between a group of sunspots, and suggested that magnetic cancellation could be the driving mechanism behind these fan-shaped jets.\\
\indent In this paper, we analyse recurrent chromospheric fan-shaped jets inside a sunspot group using high-resolution observations from GST, along with coordinated data from AIA and the Helioseismic and Magnetic Imager (HMI, \citealt{2012SoPh..275..207S}) onboard SDO. We also examine the photospheric field and magnetic topology to explore the driving mechanisms and evolutionary dynamics of these jets.
\begin{figure*}[t]
\renewcommand{\thefigure}{3}
\centering
\includegraphics[width=\textwidth]{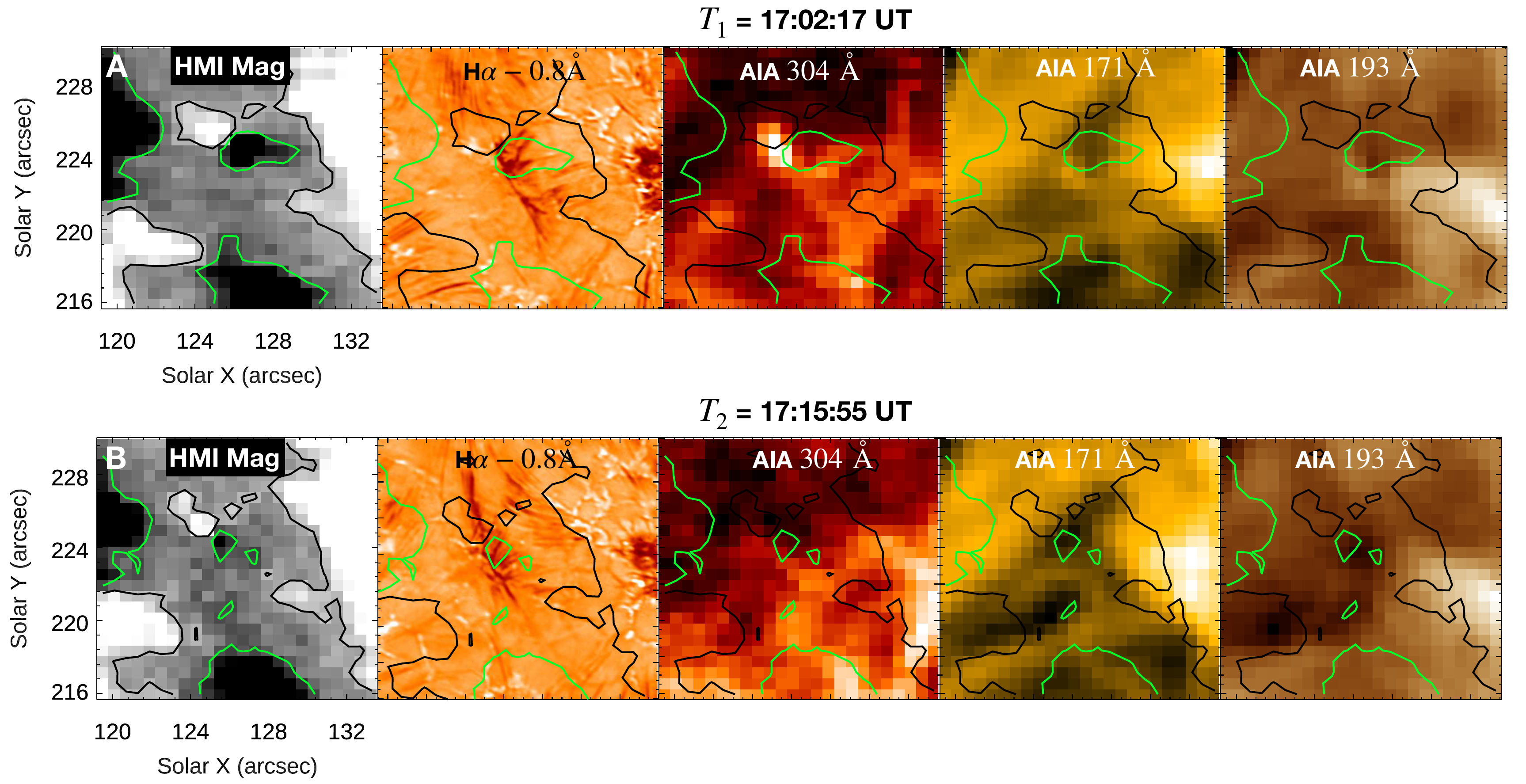}
\caption{Top and bottom panels show snapshots of the fan-shaped jets at two different time instances as seen in different AIA filters. The top panels reveal the presence of strong brightening in AIA 304 \AA\ channel and faint but less distinct brightening in AIA 171 \AA\ and AIA 193 \AA\ channels at 17:02:17 UT. The later phase images of different AIA filters (for example, in the bottom subpanels) do not show any noticeable brightening at the canceling neutral line in any of the AIA images at 17:15:55 UT.} 
\label{fig: fig3}
\end{figure*}
\section{Observations} 
We used data from the Visible Imaging Spectrometer (VIS) on the 1.6-meter GST at BBSO, taken on June 16, 2017, in the active region NOAA AR 12663. The VIS instrument captures narrowband H$\alpha$ line scan images with a single Fabry-Pérot etalon, providing a 0.07~\AA\ bandpass across the 550-700 nm range. To achieve high-cadence observations, bursts of 25 frames were taken at five positions along the H$\alpha$ line ($\pm 0.8$ \AA, $\pm 0.4$ \AA, and the line core). The spatial resolution of the H$\alpha$ images is 0$''$.029 pix$^{-1}$, and cadence is $\sim$29~s. GST data were available from 16:46:10 UT to 20:28:22 UT. All bursts were flat-field corrected and processed using speckle reconstruction to reach a diffraction-limited resolution. The images were then aligned and de-stretched to remove residual image distortion due to seeing and telescope jitter. Figure~\ref{fig: fig1} shows H$\alpha$ line wing images (H$\alpha$ - 0.8 \AA\ and H$\alpha$ - 0.4 \AA), where fan-shaped jets appear as dark elongated features that form a rosette-like structure.\\
We also used data from AIA and HMI onboard SDO. The AIA and HMI data were calibrated with the $aia_{-}prep.pro$ routine in SolarSoft (SSW). To co-align the AIA and GST data, we matched network bright points visible in both AIA 1700 \AA\ and GST H$\alpha$ - 0.8~\AA\ images. 
\section{Data Analysis and Results} 
This study investigates long-lasting recurrent fan-shaped jets occurring between a group of same-polarity sunspots by analyzing their dynamics and the evolution of the photospheric magnetic field to understand the underlying mechanisms responsible for their generation. As shown in panel B of Figure~\ref{fig: fig1}, the region's sunspots are all of the same positive polarities. Figure~\ref{fig: fig1} shows the jets as dark absorption features in H$\alpha$ - 0.8 \AA\ (panels C, D) and H$\alpha$ - 0.4 \AA\ (panel E) images. They persist for an hour and are located between a group of sunspots, one of which is a well-developed, isolated sunspot with a clear umbra and penumbra (panel A). Unlike the fan-shaped jets in LBs, which typically form within an umbra and divide it into two parts, these jets occur between distinct sunspots.\\
\indent Figure~\ref{fig: fig2} compares the temporal evolution of these jets in GST H$\alpha$ - 0.8 \AA\ images (panel A) and HMI line of sight magnetograms (panel B) at five timestamps. The blue and red contours correspond to magnetic field strengths of +20~G and -20~G, respectively. The time in panel A at the bottom is shown in minutes, starting from
\begin{figure*}[!htbp]
\renewcommand{\thefigure}{4}
\centering
\includegraphics[width=\textwidth]{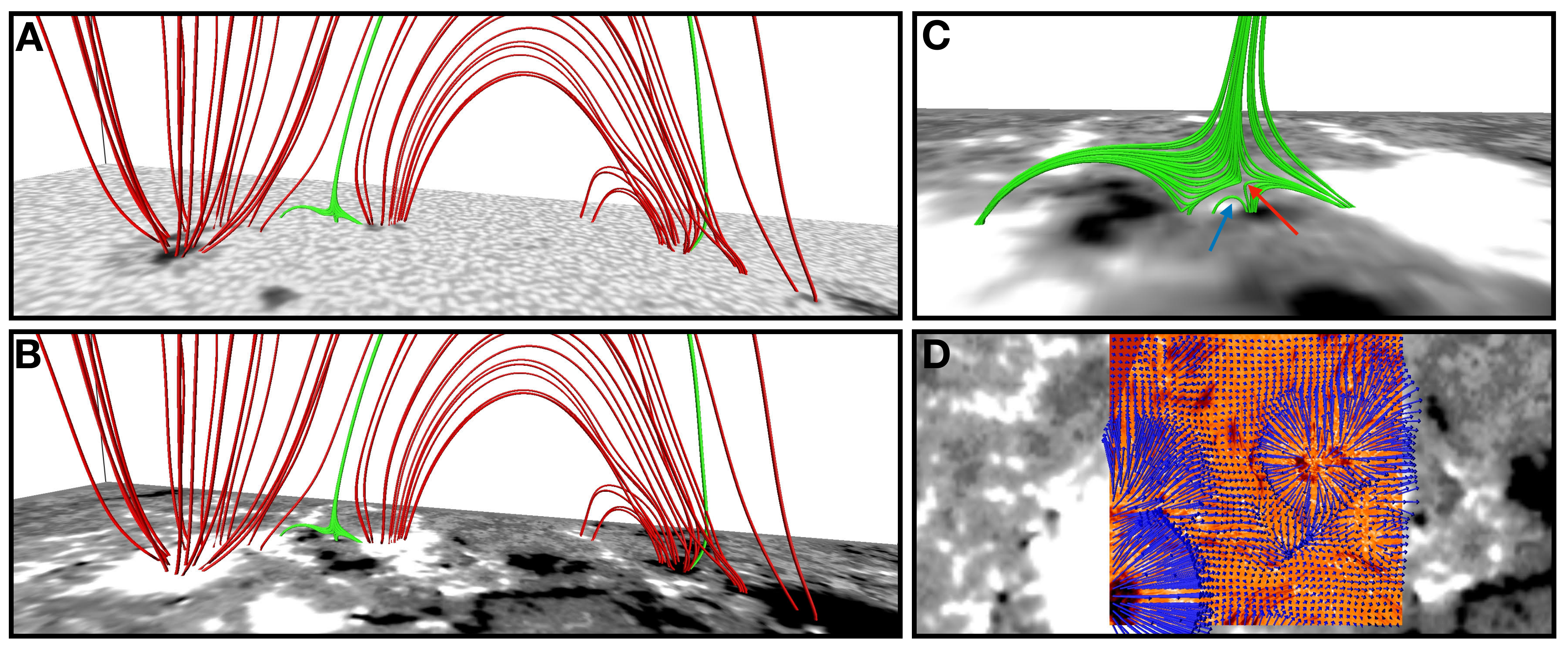}
\caption{A-B shows the NFFF extrapolated magnetic field lines of a large region at 17:00:46 UT plotted over an HMI continuum image and an HMI line-of-sight magnetogram, respectively. Panel C highlights the 3D magnetic null point structure near the location of the fan-shaped jets. Panel D shows the magnetic field strength and direction (indicated by the blue arrows) at a height of $\sim$1.25 Mm.}
\label{fig: fig4}
\end{figure*}
16:46:10 UT. The white arrows highlight the base of the dark recurrent fan-shaped jets. The comparison reveals that opposite-polarity fluxes are present at the jet base. When magnetic cancellation occurs between them, fan-shaped jets become visible in H$\alpha$ - 0.8 \AA\ images. Notably, the jet activity becomes more intense around 17:12:37 UT (approximately 26 minutes in panel D), coinciding with a significant decrease in the strength of the opposite polarity fluxes. To further examine the presence of magnetic cancellation, we analyzed the magnetic flux evolution within the regions marked by the blue and red boxes in panel C of Figure~\ref{fig: fig2}, which enclose the positive and negative flux, respectively. After carefully inspecting the magnetogram movie, we manually placed the boxes to ensure that the evolving canceling magnetic elements were fully captured during their lifetimes. To avoid contributions from the opposite polarity within the selected boxes, we calculated the magnetic flux by only including those pixels where $B_{los}>+20$ G for positive flux and $B_{los}<-20$ G for negative flux. The corresponding magnetic flux time profile is shown in panel D of Figure~\ref{fig: fig2}, with each curve representing the normalized flux from its respective region. The decreasing trend of both polarity fluxes over time confirms ongoing magnetic cancellation at the base of the recurrent fan-shaped jets. Further, we roughly estimated the area of the fan-shaped jets within the golden rectangular region marked in panel A of Figure~\ref{fig: fig2}. To identify the jet material, we apply an intensity threshold limit to isolate dark absorption features of the jet within this region. The temporal variation of the resulting jet area is shown as the golden curve in panel D of Figure~\ref{fig: fig2}. It peaks around 26 minutes, which coincides with a significant reduction in opposite polarity fluxes, and gradually decreases thereafter as the cancellation diminishes.\\
We further studied the multi-wavelength appearance of these jets across different AIA channels.  Figure~\ref{fig: fig3} shows snapshots of the fan-shaped jets at two different time instances as seen in HMI LOS magnetograms, GST H$\alpha$ - 0.8 \AA\ images, and images from the AIA 304 \AA, 171 \AA\ and 193 \AA\ channels. The first snapshot corresponds to the early phase (at 17:02:17 UT), when both polarity fluxes are present, while the second snapshot represents the later phase (at 17:15:55 UT), after a significant reduction in flux. The black and green contours in all panels of Figure~\ref{fig: fig3} mark magnetic field strengths of +20 G and -20 G, respectively. In the early phase (panel A), we observe strong brightening in AIA 304 \AA\ along with faint but less distinct brightening in AIA 171 \AA\ and 193 \AA.  However, in the later phase (panel B), no noticeable brightening is present in any AIA channel, suggesting that the observed brightening is linked to the process of magnetic flux cancellation. The strong brightening observed in the AIA 304 \AA\ channel, with hardly discernible enhancement in the AIA 171 \AA\ and 193 \AA\ channels, suggests that the reconnection process is likely occurring at lower atmospheric heights, most likely in the chromosphere.\\
\indent We further analyzed the magnetic field structure of AR 12663 using the numerical non-force-free fields (NFFFs) extrapolation method developed by \citet{2008SoPh..247...87H} and \citet{2008ApJ...679..848H}, \citet{2010JASTP..72..219H}. Figure~\ref{fig: fig4} shows a side view of the extrapolated magnetic field lines using an HMI magnetogram in the x-y plane at 17:00:46 UT as a boundary condition. These field lines are overlaid on the HMI continuum (panel A) and HMI line-of-sight magnetogram (panel B). The extrapolation domain has 512 $\times$ 192 $\times$ 192 grid points in x, y, and z, with a spatial resolution of 0$''$.6 pix$^{-1}$, matching that of AIA. In this setup, the x and y directions correspond to the horizontal and vertical dimensions of the magnetogram, while the z direction represents height extending into the corona. The field lines are represented in red, while a subset of field lines in the vicinity of the fan-shaped jets is highlighted in green for better visibility using the visualisation software VAPOR (\citealt{2019Atmos..10..488L}). These green lines represent the magnetic field projected onto the y-z plane, allowing us to clearly identify the location of the low-lying null point. A closer view of this null point configuration is provided in panel C of Figure~\ref{fig: fig4}. Their topology is consistent with a 3D magnetic null structure described by \citet{1990ApJ...350..672L}. The height of the 3D null, as inferred from panel C of Figure~\ref{fig: fig4}, is $\sim$1.25 Mm. While the location of this chromospheric null-point can vary depending on the extrapolation method or the photospheric field used as input, our focus is on the overall fan-spine structure of the 3D null and its similarity to the chromospheric jet rather than its exact location. The low-lying green field lines connect weak negative polarity regions to the surrounding positive polarity fields, forming a dome-like fan structure and an elongated spine, which are characteristic features of a 3D null. Panel D of Figure~\ref{fig: fig4} displays the vector magnetic field over the GST H$\alpha$ - 0.8 \AA\ field of view, corresponding to the region shown in panel C of Figure~\ref{fig: fig1} at 17:00:46 UT. The blue arrows represent the transverse magnetic field at a height of $\sim$1.25 Mm, where the length of each arrow is proportional to the field strength. We observe that the magnetic field lines in the region of the fan-shaped jets transition sharply from vertical to horizontal, indicating the presence of twisted or sheared fields. This behavior is also characteristic of fan-shaped jets associated with LBs. In both cases, strong magnetic shear forms between the nearly horizontal fields and the surrounding vertical umbral fields, generating a sharp current layer that facilitates magnetic reconnection (\citealt{2009ApJ...696L..66S}; \citealt{2015ApJ...811..137T}; \citealt{2018A&A...609A..14R}). This similarity suggests that, despite occurring in different magnetic environments, the fan-shaped jets between a pair of same-polarity sunspots share key aspects with LB-associated fan-shaped jets.
\begin{figure*}[!htbp]
\renewcommand{\thefigure}{5}
\centering
\includegraphics[width=\textwidth]{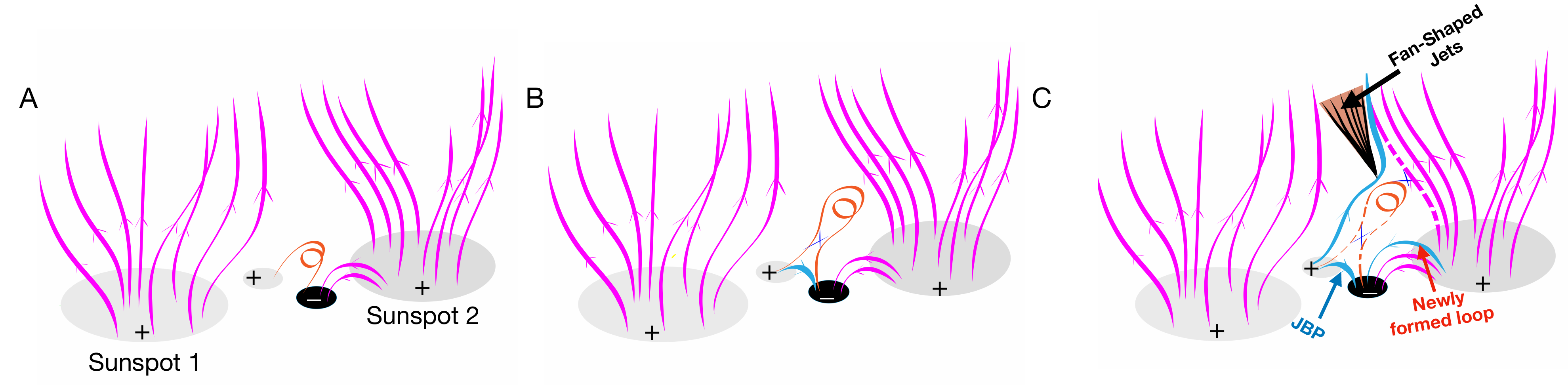}
\caption{A cartoon illustrating the proposed mechanism for fan-shaped jets between a pair of same-polarity sunspots, based on the minifilament eruption model. The magenta lines represent the overarching magnetic field, while the magenta loop lines indicate the pre-existing bipole. The orange line depicts a sheared and twisted field containing a minifilament flux rope, likely containing trapped chromospheric (minifilament) material . The crosses mark the reconnection sites, where opposite-polarity field lines interact. The small blue loop in panels B and C represents the newly reconnected field from the internal reconnection, which we suggest makes the AIA 304 \AA, 171 \AA, and 193 \AA\ brightenings at the canceling neutral line in the top panels of Figure~\ref{fig: fig3}. In panel C, the large blue loop (labeled ``Newly formed loop") results from external reconnection (upper cross), with the upper blue reconnected line facilitating plasma ejection, including the formation of the fan-shaped jets, shown as an inverted orange triangle regions.}
\label{fig: fig5}
\end{figure*}
\section{Discussion and Conclusion}
Using the joint observations of GST, AIA, and HMI, we investigated the underlying formation mechanism of recurrent chromospheric fan-shaped jets. Our analysis clearly demonstrates that magnetic flux cancellation plays a crucial role in the formation of recurrent fan-shaped jets in our observed sunspot group. The HMI LOS magnetograms reveals a gradual decrease in both the opposite-polarity fluxes at base of the jets, coinciding with the eventual disappearance of the jetting in GST H$\alpha$ - 0.8~\AA\ images (Figure~\ref{fig: fig2}). This suggests that magnetic reconnection, triggered by flux cancellation, is responsible for driving these jets. The bright fronts observed in AIA channels (discussed in section~\ref{sec:Intro}) have been interpreted in different ways—either as heating to transition region temperatures caused by shock fronts or compression (\citealt{2015MNRAS.452L..16B}) or as oscillations driven by leaked p-mode waves from the photosphere (\citealt{2015ApJ...804L..27Y}; \citealt{2017ApJ...838....2Z}; \citealt{2018ApJ...855L..19R}). In our observations here, we did not observe such fronts.  We do, however, observe strong brightenings in AIA 304 \AA\ (and weaker brightenings at the same location in AIA 171 \AA\ and 193 \AA\ channels) near the canceling neutral line early in the cancellation phase (top panels of Figure~\ref{fig: fig3}), while we do not see any such brightenings near the end of the cancellation phase (bottom panels of Figure~\ref{fig: fig3}).  This reinforces that these brightenings are linked to magnetic reconnection induced by the cancellation.\\
\indent The coronal magnetic field structure inferred using NFFFs extrapolation reveals a complex topology at the jet formation site, where a 3D magnetic null structure is present at a height of $\sim$1.25 Mm. The vector magnetogram further shows a sharp transition from vertical to horizontal fields in the vicinity of the jets, indicating strong magnetic shear. This shear might play a significant role in facilitating reconnection, similar to the process observed in LB-associated jets (\citealt{2009ApJ...696L..66S}; \citet{2011RAA....11..701J}; \citealt{2015ApJ...811..137T}; \citealt{2018A&A...609A..14R}). 
While LBs provide a localized environment with sheared horizontal fields embedded within umbral regions, our findings indicate that similar conditions can exist in intergranular lanes between sunspots. Additionally, the jets observed in our study exhibit similarities to those reported by \citet{2018ApJ...855L..19R} and \citet{2022ApJ...932...95Z}, as they are also not directly associated with LBs.\\
\indent Based on our findings however, we believe that the evidence points to the mechanism illustrated in Figure~\ref{fig: fig5} being responsible for the fan jets that we observe here. This is based on the minifilament eruption model in 2D, that was previously introduced to explain the production of coronal jets (\citealt{2015Natur.523..437S}). Panel A depicts the initial magnetic configuration, where a pair of same-polarity sunspots are separated by an intergranular lane containing weak, mixed-polarity fluxes. The magenta lines represent the overarching magnetic field, while the pre-existing larger bipole is shown in magenta loop. The smaller bipole with sheared and twisted field containing a minifilament is shown by an orange field line, and would form from magnetic reconnection resulting from canceling opposite-polarity flux patches (Figure~\ref{fig: fig2}), as is the case with coronal jets (\citealt{2016ApJ...832L...7P}). When the field containing the minifilament destabilizes, it erupts outward, guided between the larger bipole and the surrounding far-reaching umbral vertical magnetic field. As it ascends, the stretched magnetic field beneath it undergoes reconnection, a process referred to as `internal reconnection.' This forms a jet bright point (JBP) at the base of the eruption (depicted as a blue loop in panel B), corresponding to the strong brightening observed in AIA 304 \AA\ (panel A, Figure~\ref{fig: fig3}) and is also evident in our magnetic field extrapolation (marked by blue arrow in panel C, Figure~\ref{fig: fig4}). In the minifilament eruption model, brightenings typically occur along the canceling neutral line (\citealt{2018ApJ...864...68S}). Similarly, the strong brightening observed in AIA 304 \AA\ is also located at the canceling neutral line (Figure~\ref{fig: fig2},~\ref{fig: fig3}). This is further supported by our NFFFs extrapolation, which identifies an elevated 3D null point at $\sim$1.25 Mm, corresponding to the chromosphere. This suggests that the jets are primarily composed of chromospheric material, likely released (via reconnection) from an erupting mini-flux rope carrying a minifilament (\citealt{2024ApJ...974..123W}). As a result, the eruption leaves a distinct brightening in the 304 \AA\ channel, and weaker brightenings in 171 \AA\ and 193 \AA\ channels. Further upward, the outer part of the minifilament interacts with the vertical umbral field on the opposite side of the larger bipole, leading to `external reconnection.' This process forms a fan-shaped jet that propagates between far-reaching field lines (shown as orange inverted triangle region) and also adds a heated layer to the larger bipole, which is also evident in the magnetic field extrapolation (marked by red arrow in panel C, Figure~\ref{fig: fig4}). These fan-shaped jets display a rosette-like structure similar to those seen in LB fan jets. We found that fan-shaped jets cease when the minority-polarity fluxes disappears. \citet{2017ApJ...844..131P} first noted that jets persist as long as the minority-polarity flux is present, but cease when it vanishes, causing the polarity inversion line to disappear.  Similarly, \citet{2018ApJ...864...68S} observed recurrent jets in ARs that only occur during flux cancellation and stop once the minority flux is gone. This suggests that flux cancellation plays a crucial role in sustaining the jet activity.\\

Acknowledgements. We gratefully acknowledge the use of data from the Goode Solar Telescope (GST) of the Big Bear Solar Observatory (BBSO). BBSO operation is supported by the US NSF AGS 2309939 grant and the New Jersey Institute of Technology. The GST operation is partly supported by the Korea Astronomy and Space Science Institute and the Seoul National University. T.S. acknowledges that the initial part of this work was carried out during his tenure as a NASA Postdoctoral Program Fellow at NASA Marshall Space Flight Center (MSFC). A.P. would like to acknowledge the support from the Research Council of Norway through its Centres of Excellence scheme, project number 262622, and Synergy Grant number 810218 459 (ERC-2018-SyG) of the European Research Council. A.C.S. and R.L.M. received funding from the Heliophysics Division of NASA’s Science Mission Directorate through the Heliophysics Supporting Research (HSR) Program. V.Y. acknowledges the support from the NASA 80NSSC20K0025 grant and NSF AGS 2309939, 230034, 2401229, and AST 2108235.

\bibliography{fanloops}{}
\bibliographystyle{aasjournalv7}

\end{document}